\documentclass[conference]{IEEEtran}
\ifCLASSINFOpdf
  \usepackage[pdftex]{graphicx}
  \graphicspath{{../pdf/}{../jpeg/}}
  \DeclareGraphicsExtensions{.pdf,.jpeg,.png}
\else
  \usepackage[dvips]{graphicx}
  \graphicspath{{../eps/}}
\fi
\usepackage{mathtools}
\usepackage{amsmath}

\usepackage{color,soul}

\makeatletter

\makeatletter
\def\ps@IEEEtitlepagestyle{
  \def\@oddfoot{\mycopyrightnotice}
  \def\@evenfoot{}
}
\def\mycopyrightnotice{
  {\footnotesize
  \begin{minipage}{\textwidth}
  \centering
  Copyright~\copyright~2021 IEEE. Personal use of this material is permitted. However, permission to use this  \\ 
  material for any other purposes must be obtained from the IEEE by sending a request to pubs-permissions@ieee.org. \\
  This paper has been accepted at ISCAS 2021 for publication.
  \end{minipage}
  }
}

\begin{document}

\newcommand*{\red}{\textcolor{red}}
\newcommand*{\blue}{\textcolor{blue}}

%
\title{Hybrid In-memory Computing Architecture for the Training of Deep Neural Networks}




\author{\IEEEauthorblockN{Vinay Joshi\IEEEauthorrefmark{1},
Wangxin He\IEEEauthorrefmark{2},
Jae-sun Seo\IEEEauthorrefmark{2}, and
Bipin Rajendran\IEEEauthorrefmark{1} \\
Email: vinayjoshi.iitb@gmail.com, \{wangxinh,jseo28\}@ase.edu, bipin.rajendran@kcl.ac.uk}
\IEEEauthorblockA{\IEEEauthorrefmark{1}King's College London, Strand, London WC2R 2LS, United Kingdom}
\IEEEauthorblockA{\IEEEauthorrefmark{2}Arizona State University, AZ, USA}}
%


\maketitle

\begin{abstract}
The cost involved in training deep neural networks (DNNs) on von-Neumann architectures has motivated the development of novel solutions for efficient DNN training accelerators. We propose a hybrid in-memory computing (HIC) architecture for the training of DNNs on hardware accelerators that results in memory-efficient inference and outperforms baseline software accuracy in benchmark tasks. We introduce a weight representation technique that exploits both binary and multi-level phase-change memory (PCM) devices, and this leads to a memory-efficient inference accelerator. Unlike previous in-memory computing-based implementations, we use a low precision weight update accumulator that results in more memory savings. We trained the ResNet-$32$ network to classify CIFAR-$10$ images using HIC. For a comparable model size, HIC-based training outperforms baseline network, trained in floating-point $32$-bit (FP$32$) precision, by leveraging appropriate network width multiplier. Furthermore, we observe that HIC-based training results in about $50\,$\% less inference model size to achieve baseline comparable accuracy. We also show that the temporal drift in PCM devices has a negligible effect on post-training inference accuracy for extended periods (year). Finally, our simulations indicate HIC-based training naturally ensures that the number of write-erase cycles seen by the devices is a small fraction of the endurance limit of PCM, demonstrating the feasibility of this architecture for achieving hardware platforms that can learn in the field.
\end{abstract}


%

\section{Introduction}
Numerous emerging smart applications (e.g. IoT, wearables, drones, etc.) demand on-chip continuous learning, compelling the development of application-specific memories and architectures. More often these applications demand the implementation of learning algorithms for large network models in an energy-efficient manner. Conventional digital memory solutions based on SRAM or DRAM cannot address the required density/energy requirements due to large area and restrictive off-chip memory access costs. High-end expensive graphical processing units (GPUs) have been a default choice to perform DNN training. The energy and time requirement of training the state-of-the-art DNN architectures on GPUs is high \cite{gpu_cost}. This necessitates the development of more energy-/area-efficient custom hardware accelerators for performing deep learning training workloads.  A few ASIC processors have been recently reported for DNN training \cite{asic1,asic2,Yin_SSCL2020}, but based on conventional SRAM for on-chip storage, which requires a large amount of memory access with associated density and leakage power constraints. 

The multi-level storage offered by non-volatile memories such as phase-change memory (PCM) \cite{PCM}, resistive RAM (RRAM) \cite{RRAM} when combined with in-memory computing forms a basis of analog DNN training accelerators \cite{Y2018sebastianJAP,Y2017burrAPX,Joshi2020,isaac,NSI2020,prime,pipe_layer}. In-memory computing with PCM devices has shown promise in training deep neural networks (DNNs) on several cognitive tasks \cite{Y2017burrAPX,NSI2020,rpu}. However, in these solutions weight gradients are stored on off-chip storage which again inherits the inherent density/power constraints. 
In \cite{Luo_ToC_2020} a hybrid precision synapse for DNN training is proposed. In our implementation, unlike in \cite{Luo_ToC_2020}, the higher significant bits are programmed only if there is an overflow event on the lower significant bits. Furthermore, our implementation ensures that the conductance decay due to PCM device drift (similar to capacitor leakage in \cite{Luo_ToC_2020}) does not affect the network training accuracy.

Here, we propose a hybrid in-memory computing (HIC) architecture for training of DNNs on hardware accelerators and memory-efficient inference. Specifically, we propose to map a higher significant bits of the weight values using PCM devices with multi-level storage capability and the lower significant bits using PCM devices with binary-level storage capability. In the following sections, we show that this scheme can outperform baseline network trained, in floating-point $32$-bit (FP$32$) precision, by leveraging appropriate network width multiplier and requires about $50\,$\% less inference model size to achieve similar accuracy in comparison to the baseline. 


\section{DNN training using Hybrid In-memory Computing} \label{sec:hic}
In this section, we will describe the weight representation strategy on two memory arrays used in the HIC architecture. We also discuss, a set of hardware aware operations that are an essential basis of the HIC architecture.

\subsection{Weight representation strategy}
\begin{figure}
	\centering
	\includegraphics[width=0.4\textwidth]{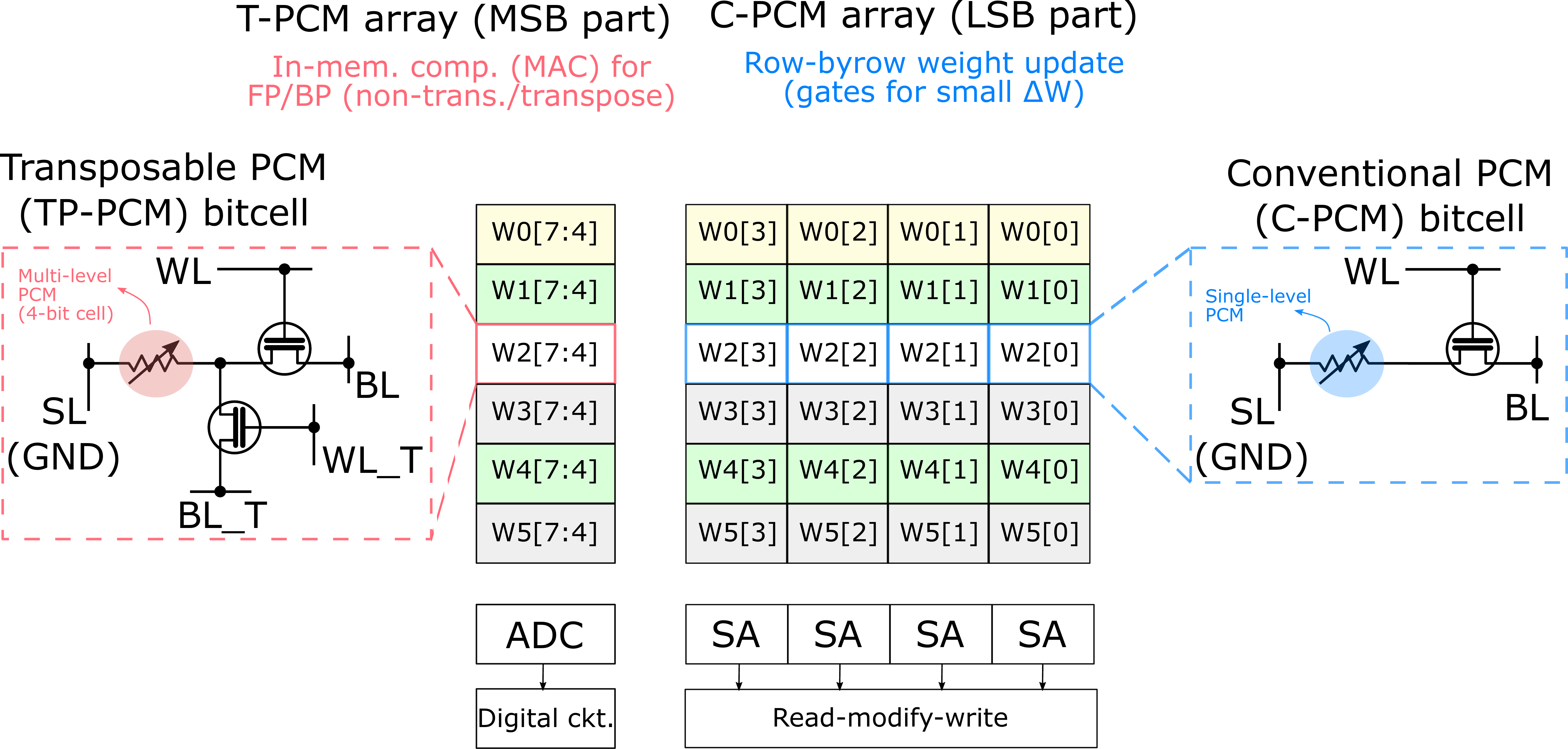}
	\caption{Hybrid in-memory computing (HIC) architecture for DNN training - the  MSB part of the synaptic weight is stored on a multi-level PCM cell that offers the equivalent of 4-bit precision while the LSB part is stored on a 7-bit memory formed by seven binary PCM devices. This choice of precision in MSB and LSB parts is found to be optimal for DNNs studied in this paper.}
\label{FIG:HIC_ARCH}
\end{figure}

An overview of the weight representation strategy in the HIC architecture is shown in Fig. \ref{FIG:HIC_ARCH}. In the HIC architecture the MSB part of the weight values is stored on an array of multi-level PCM cells denoted as MSB array. The LSB part of the weight values is stored on a memory array of binary-valued PCM devices denoted as LSB array. This hybrid design is motivated by the fact that only MSBs of weights are needed for forward/backward propagation of DNN training that targets low-precision inference, while typically weight updates are small values that mostly modify only LSBs of weights. The MSB values of weights are programmed and read using statistically accurate PCM models proposed in \cite{NSIJAP}. The LSB part of the weights is mapped using their binary representation on multiple binary-level PCM devices. Notably, the write operation on a memory location in the LSB array is performed by simply reading and flipping the binary state of the appropriate PCM device. This results in reduced write operation cost on the LSB array but also implies that some PCM devices shall drift more compared to their neighbors in the same memory location (as device drift is based on the last programming time). However, we will empirically show that this does not adversely affect the network training accuracy even for extended periods of time (year).

The MSB array is a crossbar array formed by a differential pair of multi-level PCM devices at each cross point. We perform our experiments using the PCM models proposed for DNN training in \cite{NSIJAP}. In \cite{NSIJAP}, PCM models were developed by gathering the read/write statistics from 10K PCM devices. The proposed PCM model has a strong statistical match with the obtained data. Hence, the use of this PCM model for this study is an accurate choice. The proposed PCM model consists of 4 different non-ideal components, (1) stochastic write, (2) stochastic read, (3) temporal variation in device conductance, and (4) nonlinearity of the programming curve. For the LSB array, we use a binary-level PCM model with stochastic write of high conductance state, and the read operation consists of the conductance drift coupled with stochastic read as proposed in \cite{NSIJAP}. For a binary PCM device, stochastic write is simulated by adding a zero mean fixed standard deviation Gaussian noise to the expected high state conductance value. A multi-level PCM can also be used as binary-level storage by utilizing the lowest conductance state and desired high conductance state.  

\subsection{Implementation of hybrid in-memory computing}
\begin{figure}
	\centering
	\includegraphics[width=0.35\textwidth]{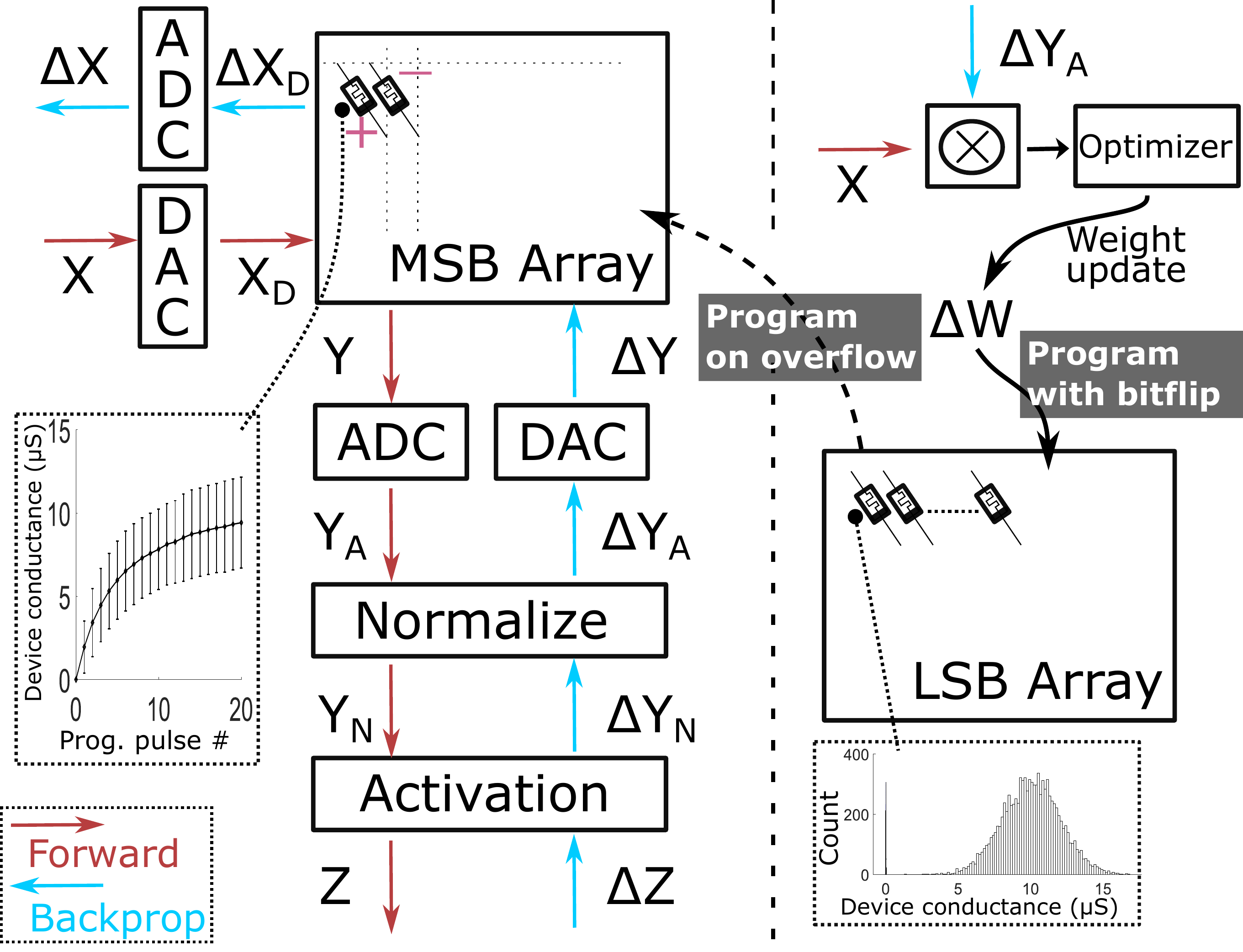}
	\caption{The overview of DNN layer operation during  HIC-based training - the MSB array is used to perform vector-matrix multiplication operations involved in the forward and backpropagation phases of DNN training, while the weight updates are quantized and accumulated in the LSB array. An overflow event in the LSB array triggers the programming of corresponding weight elements in the MSB array.}
\label{FIG:OP_LAYOUT}
\end{figure}

The HIC architecture performs vector-matrix multiplication (VMM) operation using an analog crossbar array of PCM devices. We assume that all other operations required in DNN training are performed in digital CMOS circuits. A digital to analog converter (DAC) is required to apply input voltage to read/program the analog crossbar array, and an analog to digital converter (ADC) is required to read the crossbar array output. A convolution operation is essentially a matrix-matrix multiplication \cite{cnnXbarMapping}. Hence it can be implemented on a crossbar as a vector-matrix multiplication. A recurrent layer usually performs vector-matrix multiplication followed by some elementwise non-linear activation functions \cite{lstm}. Hence, the operation layout, shown in Fig. \ref{FIG:OP_LAYOUT}, is valid for three types of commonly used DNN layers, namely fully-connected, convolution, and recurrent layer. 

Fig. \ref{FIG:OP_LAYOUT} shows a typical operation layout of a DNN layer in the HIC-based training. A transposable crossbar array performs VMM required in forward and backpropagation phases of DNN training. A digital to analog converter (DAC) applies an input (activation or error gradient) to the crossbar array. An analog to digital converter reads the output current of the crossbar array. A network-specific normalization function such as batch normalization \cite{batchnorm} or group normalization \cite{groupnorm} is computed on ADC output. An activation function such as ReLU, Sigmoid, or Tanh is computed on the normalization layer output. The gradients are backpropagated by applying voltages proportional to the error gradients on the columns of the transposable crossbar array. During the weight update phase, an outer product is computed on the input $X$ and output error gradients $\Delta Y_{A}$ to obtain weight gradients $\Delta W$. The weight gradients are quantized and used to update the LSB array. The values in the LSB array are updated by simply flipping the binary states of the devices if required. In case there is an overflow on the LSB array, the MSB array is programmed with a corresponding update value. There are no other specific programming events on the MSB array.

\section{Numerical validation} \label{sec:numerical_validation}
In this section, we discuss the experimental setup used for our simulations and the results from four different studies.
\subsection{Experimental setup and network details}
We have chosen ResNet-$32$ network for all our evaluations as they are representative of large networks used in the machine learning research \cite{resnet}. The ResNet-$32$ network has a total of $33$ convolution layers and $1$ fully-connected layer for classification. It has about $470\,$K trainable parameters and all the weights and updates are stored on PCM-based memory arrays. All the $34$ layers of the ResNet-$32$ network are trained using HIC architecture. The network is trained to classify the images from the CIFAR-$10$ dataset \cite{cifar}. 

The scheme used to program PCM devices on the MSB array can only increment the device conductance. Application of several programming pulses to the devices in a differential pair can result in saturation of their conductance level. After every $10$ batches of training, we perform a refresh operation on the PCM devices in the MSB array to avoid saturation of device conductance \cite{Boybat2018}. A differential pair of PCM devices used in the MSB array offers an equivalent precision of approximately $4$-bits \cite{pcm4-bit}. In the LSB array, we use seven PCM devices for $7$-bit signed fixed-point representation. All the DACs and ADCs have $8$-bit precision\red, as they have been reported to be design points with good trade-off between precision and energy consumption \cite{Rekhi2019}. We use the same hyperparameter setting for the baseline network and preprocessing of the CIFAR-10 images as given in \cite{resnet}. The hyperparameter setting in HIC implementation is the same as \cite{resnet} except that the learning rate is 0.05 with a decay factor of 0.45 and a batch size of 100. All our simulations are performed using TensorFlow \cite{TF}. 

\subsection{Effect of individual non-idealities}
\begin{figure}
	\centering
	\includegraphics[width=0.35\textwidth]{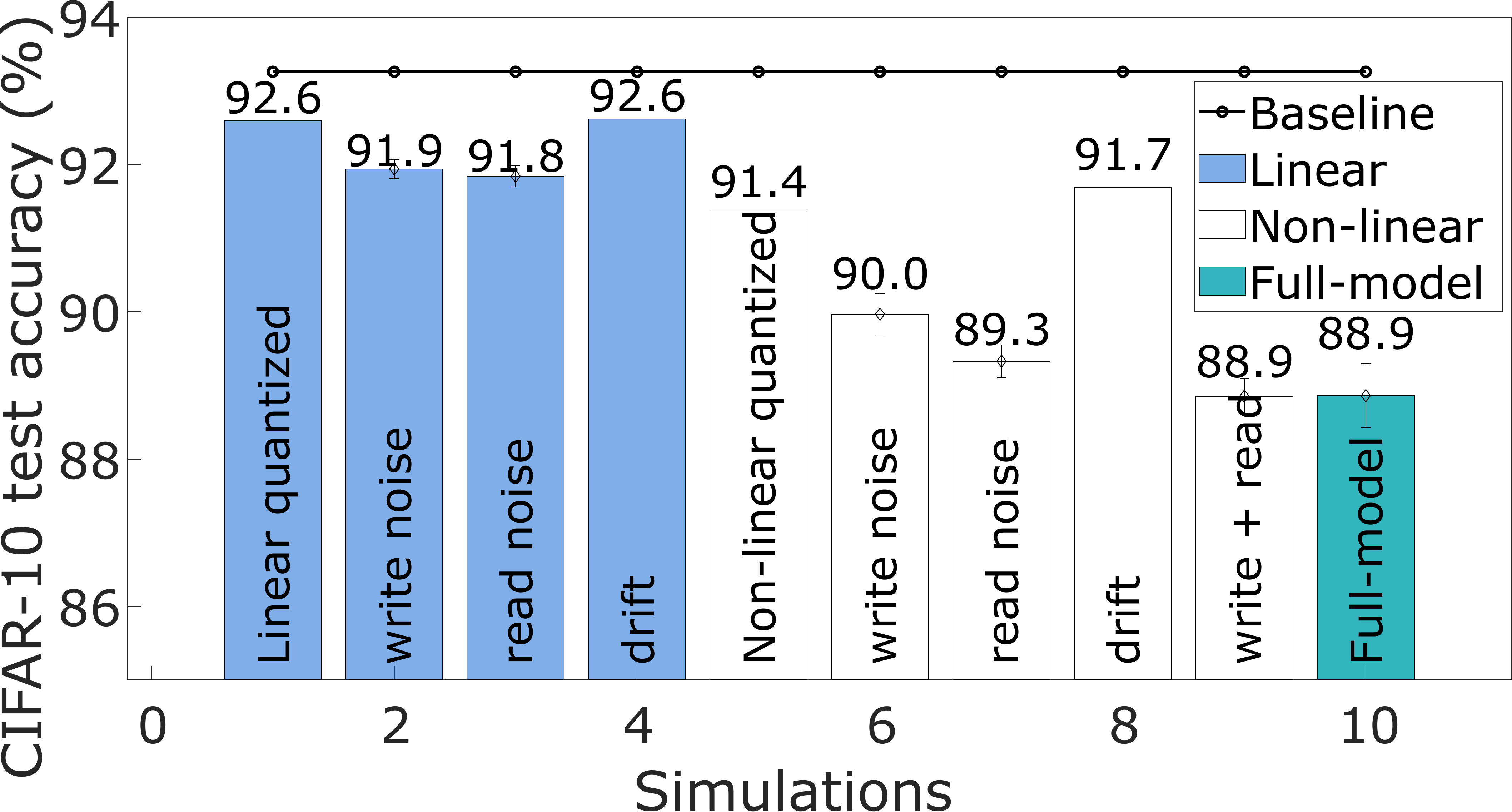}
	\caption{The   impact of different non-ideal aspects of the PCM on HIC-based training accuracy. PCM device non-linearity, as well as stochasticity associated with write and read operations results in a notable drop in the training accuracy   while the temporal drift in the device conductance results in accuracy improvement as it has the same effect on DNN training as that of the weight decay regularization. The results are based on the  average of five distinct training runs.}
	
\label{FIG:ABLATION}
\end{figure}

Our simulations are based on the PCM model proposed in \cite{NSIJAP}. This PCM model is composed of four non-ideal components, (1) stochastic write, (2) stochastic read, (3) temporal variation in the device conductance, and (4) nonlinear programming curve. We study the effect of each non-ideal component on the network training accuracy by performing the ablation of the PCM model. As shown in Fig. \ref{FIG:ABLATION}, light blue bars indicate network trained with linear PCM model and at most one other non-ideal component (see the text on bars). The colorless bar indicates a network trained with the nonlinear PCM model and at most two other non-ideal components. The right-most colored bar (denoted as "Full-model") indicates network trained with a PCM model including all the non-ideal components.

We observe that the nonlinearity of the programming curve causes a notable drop in the network training accuracy compared to the linear PCM model. This is attributed to the fact that the expected conductance increment in the PCM device is an inverse function of the number of applied programming pulses \cite{NSIJAP}. The stochastic read and write of the PCM model have a stronger negative effect on the network training accuracy due to the the large amount of write noise present in the PCM programming process. Interestingly, the implementation that incorporates device conductance drift achieves higher accuracy in comparison to other nonideal components. This accuracy improvement can be attributed to the fact that the device drift is similar to the weight decay regularization technique commonly used in deep learning \cite{wd1,wd2}. Like weight decay regularization, PCM conductance drift causes higher decay in the weight values that do not contribute to the network training and weights that are regularly updated (or trained) are subjected to smaller drift in their conductance value. 

Although the network training accuracy in the full-model case including all the non-ideal components is about 4.4\% lower than the baseline network trained in FP32 precision, note that HIC is utilizing only 9-PCM devices. 
We show in the next section that HIC-based training is more memory-efficient and it achieves better accuracy compared to the baseline.


\subsection{Effect of the model size on network accuracy}
\begin{figure}
	\centering
	\includegraphics[width=0.35\textwidth]{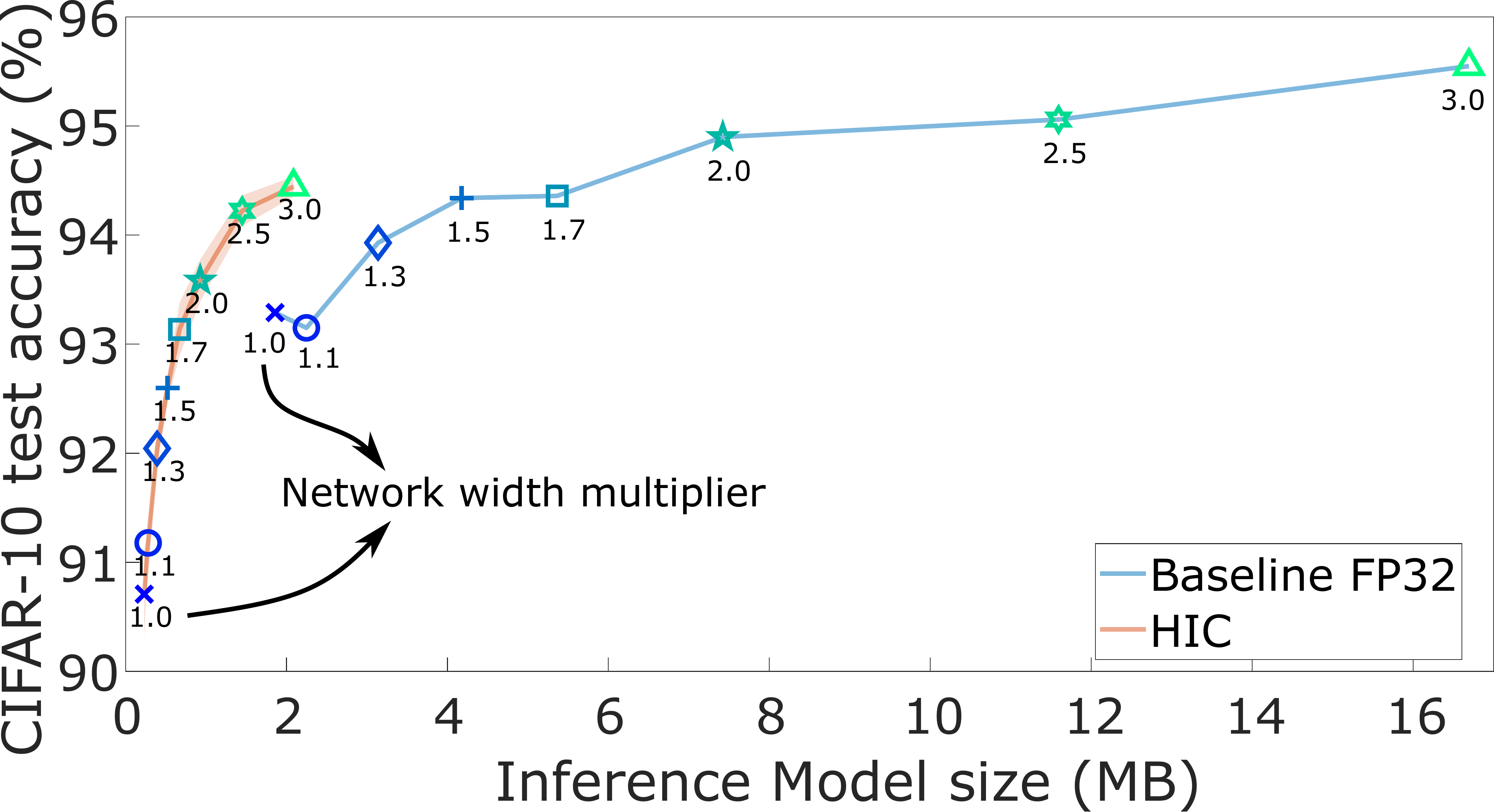}
	\caption{With network width multiplication, HIC-based training of  ResNet-32 network outperforms the FP32 software baseline for comparable inference model size and  also shows better accuracy improvement with increase in model size. Note that HIC achieves comparable accuracy to the baseline with about $50\,$\% less inference model size. The  network width multiplier (values near markers)  compensates for the loss in  accuracy due to non-idealities of the PCM devices. The results are based on the  average of five distinct training runs.}
	\label{FIG:MODEL_SIZE}
\end{figure}
Now, we discuss the network training accuracy as a function of the model size required for inference. We increase the network size by increasing the number of neurons in each layer by a desired network width multiplier (see values near markers in Fig. \ref{FIG:MODEL_SIZE}) as proposed in \cite{mobilenet}. We observe that the increase in network width compensates for the loss in training accuracy due to non-ideal PCM devices. Fig. \ref{FIG:MODEL_SIZE} shows the network training accuracy as a function of the inference model size. The inference model size is the amount of memory required to store weights during the inference phase. The HIC-based implementation requires approximately $4$-bits for storing weights and $32$-bits for the baseline. Like markers indicate the same network architecture, in other words, the same number of neurons in each layer. The model size for HIC is lower because of the low precision representation of weights. 

We observe that the accuracy improvement in HIC-based training per additional neurons/model size is better compared to the baseline. Furthermore, for a comparable inference model size, HIC-based training achieves at least 1\% better accuracy than the baseline. Notably, for a comparable network accuracy, HIC-based training requires about $50\,$\% less inference model size. This suggests that HIC-based training results in memory-efficient inference computation on the hardware without compromising on accuracy compared to the baseline.

\subsection{Effect of Device Drift on Post-training Inference Accuracy}
\begin{figure}
	\centering
	\includegraphics[width=0.35\textwidth]{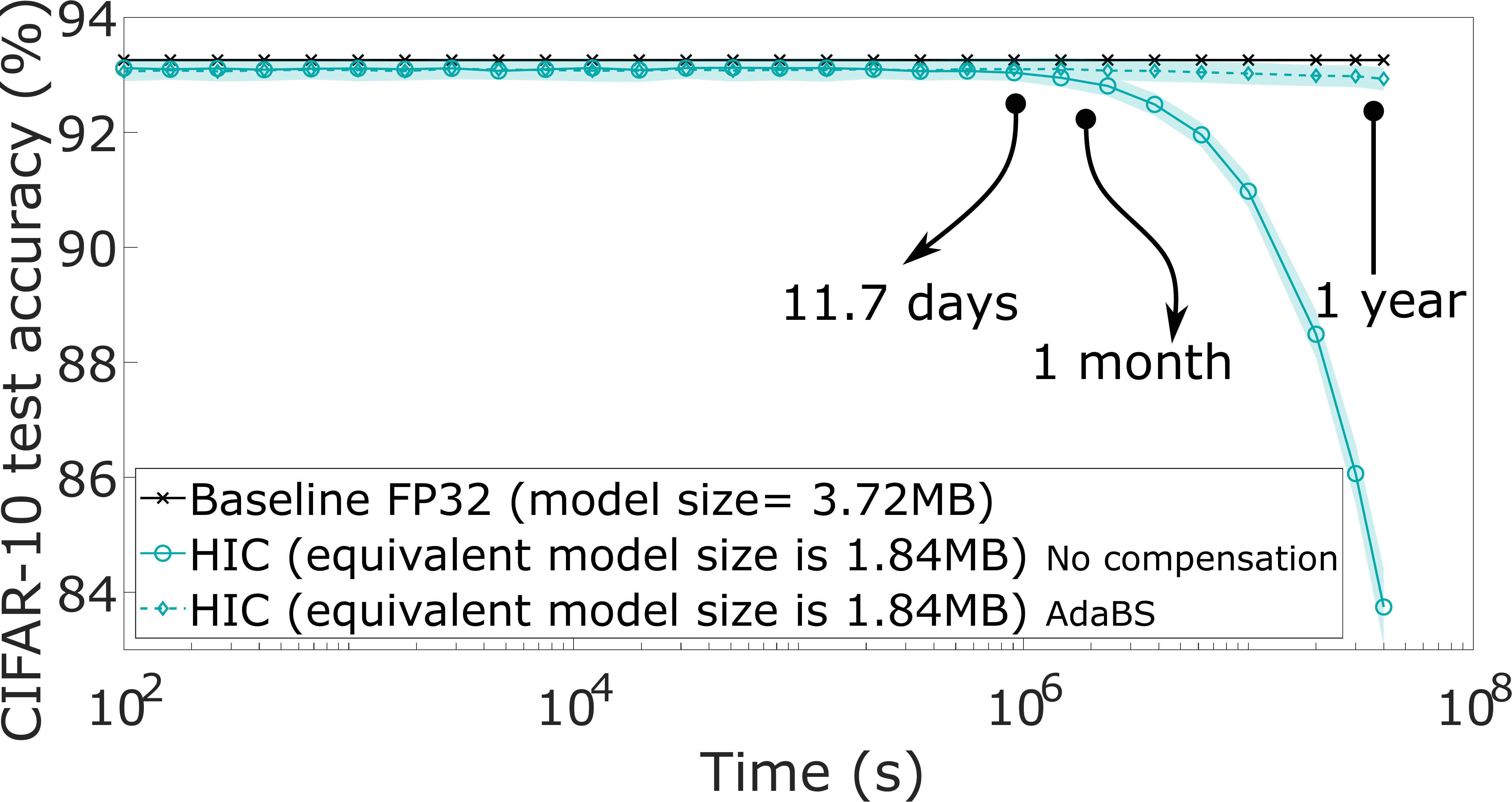}
	\caption{Post-training inference accuracy estimate of  ResNet-$32$ network  caused by the temporal drift in PCM devices - for a period of $10^{6}\, s$ (11.7\,days) there is no observable drop in the inference accuracy. The AdaBS compensation technique \cite{Joshi2020} is shown to be effective in compensating for  weight degradation due to PCM drift,  resulting in maintaining the inference accuracy   close to the baseline  for over a year. The plots shown are averaged over 10 distinct training runs and 10 distinct inference runs per training (total of 100 runs).}
	\label{FIG:ACCURACY_RETENTION}
\end{figure}
Now, we discuss the effect of PCM conductance drift on the inference accuracy computed as a function time after network training.  As shown in Fig. \ref{FIG:ACCURACY_RETENTION}, we compute the inference on a version of the HIC network that uses a width multiplier of $1.7$. 
In these simulations, we initially train the network for $205$ epochs and incorporate the drift in the PCM devices which causes the weights to degrade throughout training. We perform the inference on the trained network from $100\,$s to $4\times10^{7}\,$s causing further degradation in the weight value. 

In \cite{Joshi2020}, AdaBS technique was proposed to compensate for the weight decay caused by the PCM conductance drift. The AdaBS technique infrequently performs a calibration phase that recomputes the global mean and variance of all the batch normalization layers in the network. The calibration phase requires about $5$\% of the images in the training set. 

The inference accuracy of HIC is unaffected for the time duration of about $10^{6}\,$s when no compensation is applied to the weights. After $10^{6}\,$s, there is more adverse impact of the drift on the inference accuracy. The AdaBS compensation technique helps to recover the degradation in the network accuracy. For a year long simulation the inference accuracy of HIC drops by just $0.12\,$\% compared to the training accuracy at $100\,$s when AdaBS compensation is applied on the weights. The drop in the inference accuracy is $9.37\,$\% when no compensation is applied and AdaBS helps to recover such a significant drop in the network accuracy. Note that using AdaBS compensation no significant gain in the inference accuracy is observed till $10^6\,$s but afterwards AdaBS plays a crucial role in maintaining the inference accuracy close to the baseline.

\subsection{Write-Erase Cycle Estimation}
\begin{figure}
	\centering
	\includegraphics[width=0.4\textwidth]{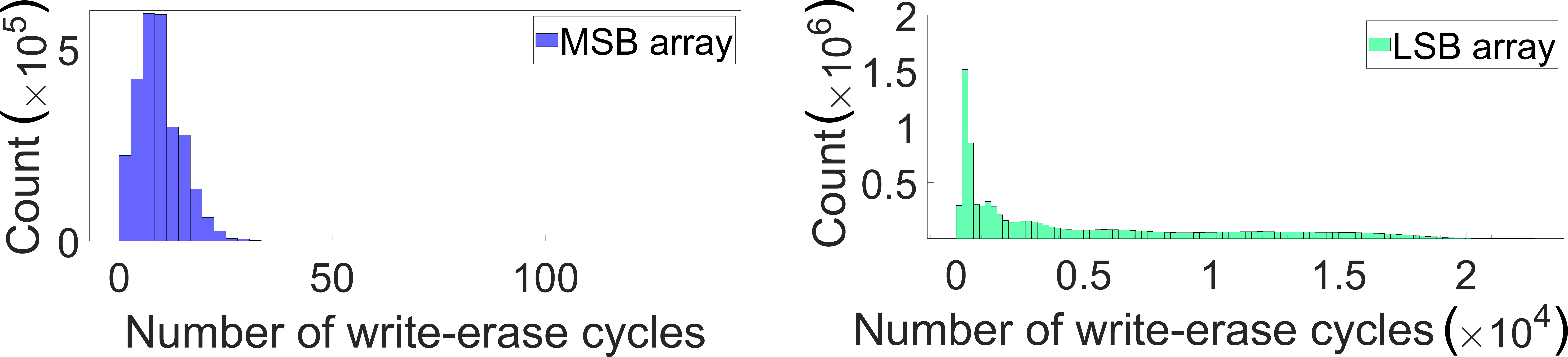}
	\caption{The number of write-erase cycles applied to the PCM devices on the MSB or LSB array during the training of ResNet-$32$ network is  well within the reported endurance limit of   PCM devices ($10^{8}$).}
	\label{FIG:ENDURANCE}
\end{figure}

During our training simulations we also tracked the number of write-erase cycles that were applied to the devices on the MSB or LSB array throughout the training. 
Following the definition in \cite{pcm_endurance}, we define a write-erase cycle as a sequence of at most $10$ SET pulses followed by a RESET pulse. Fig. \ref{FIG:ENDURANCE} shows the distribution of write-erase cycles applied on all devices during one full training of the ResNet-$32$ network. The number of write-erase cycles applied to a PCM device in the HIC-based training of the ResNet-$32$ network is less than $150$ for the MSB array and less than $20\,$K for the LSB array. A PCM device endurance is of the order of $10^{8}$ \cite{pcm_endurance} and the number of write-erase cycles seen by any device in HIC implementation is a small fraction of PCM endurance.


\section{Conclusion}
\label{SEC:Conclusion}
We proposed hybrid in-memory computing (HIC) architecture for memory-efficient training of deep neural networks on hardware accelerators. Based on the simulations demonstrating training of ResNet-32 network on CIFAR-10 dataset, the HIC architecture shows promising memory savings and higher training accuracy compared to the baseline network trained in floating-point 32-bit precision. Specifically, the HIC implementation outperformed baseline software accuracy by at least $1\,$\% while still using a comparable amount of inference model size by leveraging a network width multiplier. Interestingly, the HIC implementation achieved similar accuracy to that of baseline but with $50\,$\% less inference model size by leveraging a network width multiplier. With a suitable PCM device drift compensation technique, we showed that post-training inference accuracy suffers a negligible drop. Finally, the HIC-based training incurred write-erase cycles that are a small fraction of PCM endurance demonstrating the usefulness of this architecture for the memory-constrained deep learning applications such as edge computing, IoT, wearable technology.

\section*{Acknowledgment}
This work was in part supported by NSF grants 1652866
and 1715443, SRC AIHW program, and C-BRIC,
one of six centers in JUMP, a SRC program sponsored by
DARPA.

\bibliographystyle{IEEEtran}
\bibliography{ref}
\end{document}